# Ab initio study of Li, Mg and Al insertion into rutile VO$_2$: Fast diffusion and enhanced voltages

*Vadym V. Kulish, Daniel Koch and Sergei Manzhos$^a$\**

$^a$ Department of Mechanical Engineering, National University of Singapore, 21 Lower Kent Ridge Rd, Singapore 119077

*E-mail: mpemanzh@nus.edu.sg

**Abstract.** Vanadium oxides are among the most promising materials that can be used as electrodes in rechargeable metal-ion batteries. In this work, we systematically investigate thermodynamic, electronic and kinetic properties associated with the insertion of Li, Mg and Al atoms in rutile VO$_2$. Using first-principles calculations, we systematically study the structural evolution and voltage curves of Li$_x$VO$_2$, Mg$_x$VO$_2$ and Al$_x$VO$_2$ (0<x<1) compounds. The calculated lithium intercalation voltage starts at 3.50 V for single-atom insertion and decreases to 2.23 V for full lithiation, to the LiVO$_2$ compound, which agrees well with experimental results. The Mg insertion features a plateau about 1.6 V up to Mg$_{0.5}$VO$_2$ and then another plateau-like region at around 0.5 V up to Mg$_1$VO$_2$. The predicted voltage curve for Al insertion starts at 1.98 V, followed by two plateaus at 1.48 V and 1.17 V. The diffusion barrier of Li, Mg and Al in the tunnel structure of VO$_2$ is 0.06, 0.33 and 0.50 eV, respectively. The demonstrated excellent Li, Mg and Al mobility, high structural stability and high specific capacity suggest a promising potential of rutile VO$_2$ electrodes especially for multivalent batteries.

## 1. Introduction

Wide-spread use of hybrid and all-electric vehicles and grid integration of renewable sources require reliable and affordable energy storage.[1,2] Rechargeable batteries, such as Li-ion batteries, provide the benefits of high energy density, long cycle life, and desirable power performance.[3-5] However, the limited resources of Li in the Earth's crust limit the future prospects of Li-ion batteries, making the use of lithium expensive in the long term. Therefore, the development of alternative battery concepts is essential. Among non-Li technologies, multivalent Mg and Al-ion batteries can provide the benefits of low cost, elemental abundance, and environment-friendly chemistry.[6-10] Despite the progress in Li-ion technology, identification of suitable cathode materials for post-Li batteries is still a challenge.[11-13] The intercalation of Mg and Al into a host crystal structure is typically accompanied by a strong Coulombic effect induced by the two/three electrons carried by the Mg/Al cation, resulting in unfavorable insertion energetics and slow diffusion.

Vanadium oxides (VO) are attracting research attention since they are able to operate as electrodes for the most actively studied types of metal-ion batteries.[14-21] For instance, completely reversible Li intercalation and deintercalation were achieved in V$_2$O$_5$ nanowires and nanoparticles.[16,17] Layered V$_2$O$_5$ has shown Mg storage capacities of 150 mAh/g at 2.3-2.6 V vs Mg/Mg$^{2+}$.[22] Al insertion in V$_2$O$_5$ has been demonstrated as well, although at a lower voltage of about 1 V (vs Al/Al$^{3+}$).[23,24] Specific capacity of 134-192 mAh/g at a current density of 500 mA/g has been achieved during Li insertion in V$_6$O$_{13}$ cathode.[21,25] Although V$_2$O$_5$ has been the most studied VO cathode material,[26-32] its practical application is hindered by structural instability and vanadium dissolution.[33] Thus, exploration of alternative, non-layered VO materials is highly necessary. In a recent experimental study, Park et al. reported a high Li storage capacity (320 mAh/g) and good stability in a $VO_2$ compound with edge-sharing $VO_6$ octahedra bilayers.[34, 35] However, to the best of our knowledge, there have been only limited reports on multivalent (Mg and Al) insertion in $VO_2$ compounds.[36]

In this work, we systematically investigate thermodynamic, electronic and kinetic properties associated with the insertion of Li, Mg and Al atoms in rutile $VO_2$(R). In our previous study,[37] we performed a computational pre-screening of multiple VO phases and identified the $VO_2$(R) as a promising electrode material based on single-atom Li/Mg/Al insertion. While the initial results are very promising, there are still fundamental questions that have to be answered in order to evaluate the potential of $VO_2$(R) electrodes. For instance, how does insertion energetics change with Li, Mg and Al concentration? What is the maximum achievable specific capacity and typical intercalation voltage? What is the preferred ordering of Li, Mg and Al atoms at higher concentrations? This work addresses these questions. Specifically, we use ab initio methods to study the structural evolution and voltage-capacity curves of $Li_xVO_2$, $Mg_xVO_2$ and $Al_xVO_2$ (0<x<1) compounds. Based on the computed convex hull, we identify the main ground states which exist during Li, Mg and Al insertion. While lithiation of rutile $VO_2$ has been reported experimentally,[38] there are no reports of magnesiation and aluminization of this phase. We therefore use the computed lithiation curve as a benchmark test of the model and predict voltage curves for $Mg_xVO_2$ and $Al_xVO_2$. We analyse the mechanism of Li, Mg, and Al insertion and establish a significant contribution from oxygen redox which enables achieving higher capacities than what would follow based on the simple rule of 1 valence electron being accommodated by 1 vanadium atom. The computed low Li, Mg and Al diffusion barriers, high structural stability and high specific capacity suggest a promising potential of rutile $VO_2$ electrodes for multivalent batteries.

## 2. Computational Methods

The calculations were performed using density functional theory (DFT), as implemented in the VASP 5.3 package.[39] The core electrons were treated within the projector augmented wave (PAW) method.[40, 41] The following valence electron configurations were used: Li ($2s^1$), Mg ($2p^6 3s^2$), Al ($3s^2 3p^1$), V ($3p^6 3d^4 4s^1$) and O ($2s^2 2p^4$). The plane-wave cutoff was 500 eV. Exchange-correlation effects were described with the generalized gradient approximation and the PBEsol functional.[42] To describe accurately the $d$ orbitals of the transition metal species, we used a DFT+U approximation of Dudarev et al.[43] Following the work by Ceder et al.,[44] we set $U_{eff}$ to 3.1 eV for vanadium. The Gaussian smearing with a smearing factor of 0.1 eV was used in all calculations. The optimized structures were obtained by relaxing all atomic positions and lattice vectors using the conjugate gradient algorithm until all forces are smaller than 0.02 eV Å$^{-1}$. Activation barriers for diffusion are calculated using the nudged elastic band (NEB) method.[45] The NEB calculations were performed using 3-6 intermediate images, and the initial guess of the migration pathway was generated by linear interpolation between the initial and final points of the diffusion path. Spin polarization is included, and all calculations were performed for the ferromagnetic (FM) spin ordering, which is a ground-state magnetic state of $VO_2$(R).[46-48] We have also found that the total energies of the Li-inserted phases with FM ordering are lower than those with AFM ordering (Table S1 in the Supporting information). Thus, all presented results are according to FM ordering. With this computational setup, experimental lattice parameters of different phases and different stoichiometries of vanadium oxide can be accurately reproduced.[37] A simulation cell with 2×2×4 $VO_2$ units was used, of size about 9.12×9.12×11.40 Å. The Brillouin zone was sampled with a 3×3×3 k-points set which provided converged energies to within 0.005 eV. A larger k-points set was used to analyse the electronic structure.

## 3. Results and Discussion

### 3.1. Crystal structure of $VO_2$ host

Recently, vanadium dioxide ($VO_2$) has attracted much research attention due to the rich interplay between crystal structure, atomic distortions and electronic properties.[49-53] It was reported that $VO_2$ can occur in two main phases (Fig. 1) – rutile (R, space group P42/mnm) and monoclinic (M, space group P21/c), which are very close in the VO phase diagram.[54] The $VO_2$(R) phase has a regular rutile structure, similar to other metal oxides ($MnO_2$, $TiO_2$); meanwhile, in the $VO_2$ (M), the V atoms form V-V dimers and tilt with respect to the rutile c axis. The $VO_2$(R) is metallic, while $VO_2$(M) has an optical bandgap of about 600 meV.[46-48] The mechanism of the R-M phase transition[55] is quite complex and still remains under debate. There has been no agreement on whether the transition is mainly

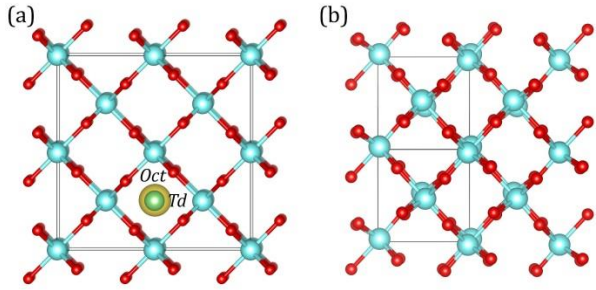

**Fig. 1** Crystal structures of (a) rutile (R) and (b) monoclinic (M) phases of $VO_2$. The typical Li, Mg and Al insertion sites in $VO_2$(R) supercell are shown. The V, O and M (M=Li, Mg, Al) atoms are shown in cyan, red and yellow/green colors, respectively

governed by the structural changes (Peierls distortion), or modification of electron-electron correlation (as in Mott insulators). It is important to note that the metal-insulator transition in $VO_2$ is quite challenging to describe by theoretical methods. For instance, the DFT+DMFT[56] and DFT+GW[47] calculations indicate that how the correlation is treated changes the calculated properties dramatically.

Due to the close proximity of $VO_2$(R) and $VO_2$(M) in the VO phase diagram, we considered both phases in this study. It is also important to check whether any phase transitions will take place during Li/Mg/Al insertion. We have calculated the total energies of $Li_xVO_2$(R) and $Li_xVO_2$(M) at $x$=0.03 and $x$=0.25 Li concentration. Our results indicate that $VO_2$(R) is more stable at both Li concentrations (Table S2). Note that the obtained difference in energy between $VO_2$(R) and $VO_2$(M) upon Li insertion is too significant for the energetic preference of the phases to be changed by vibrations. Our results are consistent with the observations that hydrogen doping in $VO_2$ nanowires stabilizes metallic rutile structure.[57, 58] Another study showed that interstitial B dopant atoms hinder the rutile-to-monoclinic transition by impeding the dimerization of V-V chains.[53] Moreover, the x-ray diffraction (XRD) measurements[38] identified the presence of $VO_2$(R) phase during Li-ion battery cycling, consistent with our calculations.

### 3.2. Li, Mg and Al insertion sites and their energetics

We first summarize the results for single-atom Li/Mg/Al insertion.[37] The calculated lattice parameters for rutile $VO_2$(R) are $a$=$b$=4.55 Å and $c$=2.76 Å, in good agreement with the experimental values ($a$=$b$=4.55 Å, $c$=2.86 Å).[59]

Each V atom is coordinated by six O atoms, forming $VO_6$ octahedrons with V-O bond lengths of 1.92 Å. The neighbouring octahedra share edges, forming tunnel-like structure along the c axis. These tunnels provide a large accessible space for Li, Mg and Al atoms. There are two distinct insertion sites for metal ions in the $VO_2$(R) structure: octahedral *(Oct)*, where Li/Mg/Al atom is bonded to six S atoms, and tetrahedral *(Td)*, where Li/Mg/Al is bonded to four S atoms. Insertion at both sites is energetically favourable with binding energies that are larger than bulk cohesive energies of Li, Mg and Al.[60] This suggests that intercalation can be successfully realized without forming detrimental metal clusters. There is a strong preference for insertion at the *Oct* site for all atoms, while the *Td* site is energetically metastable and may be potentially occupied at larger metal-ion concentrations. The computed difference in energy between the *Oct* and the *Td* sites is 0.08, 0.14 and 0.23 eV for Li, Mg and Al, respectively. Interestingly, the estimated voltages for single-atom Mg and Al insertion, corresponding to 20.18 and 30.27 mAh/g, respectively, are relatively high (2.10 and 1.99 V, respectively). Considering that the low Al insertion voltages have been a major bottleneck for the energy density in Al-ion batteries, these results suggest that $VO_2$(R) has a good potential as electrode material for multivalent batteries. In this work, we therefore explore the rutile phase further by computing the voltage-capacity curves.

### 3.3. Concentrations effects and voltage-composition curve

Our charge analysis shows that Li, Mg and Al atoms in $VO_2$(R) are almost completely ionized. Therefore, the Li-Li, Mg-Mg and Al-Al electrostatic interactions are expected to be significant. To evaluate this effect, we calculated the energy increase associated with bringing two isolated Li (Mg, Al) ions near each other within the same tunnel. As shown in Fig. 2, the computed insertion energy per Li (Mg, Al) atom shows a clear dependence on the distance between the Li (Mg, Al) atoms. We can also notice a large difference in the slope of insertion energy curves in Fig. 2. As dopant-dopant distance becomes smaller, the Li, Mg and Al insertion energy drops by 0.14, 0.29 and 0.56 eV, respectively. The larger ionic charges of $Mg^{2+}$ and $Al^{3+}$ lead to a stronger electrostatic repulsion. The computed insertion energies show that the simultaneous occupation of adjacent sites is energetically unfavourable. Therefore, Li, Mg and Al clustering is not expected in $VO_2$(R).

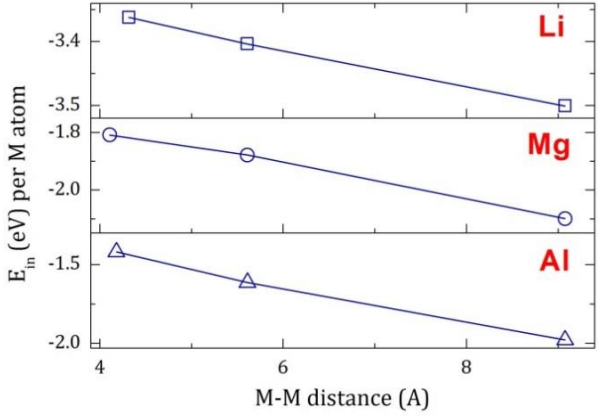

**Fig. 2** Insertion energies per M atom of two M (M=Li, Mg, Al) atoms versus M-M (Li-Li, Mg-Mg and Al-Al) distance in VO$_2$(R). The energies are calculated relative to Li, Mg and Al-bulk

Next, we calculated the total energy of different M-vacancy arrangements in the VO$_2$(R) host for concentrations ranging from $x = 0$ to 1 in M$_x$VO$_2$. For each concentration, we considered various configurations with Li, Mg and Al ions residing in either octahedral or tetrahedral sites. The inserted atoms were initially distributed as far away from each other as possible in order to minimize their self-interaction. All structures were then fully optimized. We define the formation energy ($E_{form}$) for a given intermediate composition $x$ as[61, 62]

$$E_{form} = E(M_xVO_2) - xE(MVO_2) - (1-x)E(VO_2) \quad (1)$$

where $E(M_xVO_2)$ is the total energy of M$_x$VO$_2$ per formula unit, $E(MVO_2)$ and $E(VO_2)$ are the total energies of the fully-intercalated and empty host. According to this definition, the $E_{form}$ of M$_x$VO$_2$ reflects the relative phase stability of that structure with respect to phase separation into a fraction $x$ of M$_x$VO$_2$ and a fraction (1-$x$) of VO$_2$.[61, 62] The convex hulls of formation energies of the Li$_x$VO$_2$, Mg$_x$VO$_2$ and Al$_x$VO$_2$ structures are plotted in Fig. 3b. The structures that are thermodynamically stable compared with the reference phases lie on the lower convex hull of $E_{form}$ versus composition $x$ (solid line in Fig. 3b). We notice that all considered Li$_x$VO$_2$, Mg$_x$VO$_2$ and Al$_x$VO$_2$ structures have negative formation energy, indicating that they are energetically stable. The depth of computed convex hulls increases going from Li to Mg to Al. As suggested by Radin et al, a deeper convex hull (more negative value) corresponds to stronger intercalant-intercalant repulsion.[63] From our calculations, insertion at the *Oct* sites in VO$_2$(R) is favoured over the *Td* sites for all concentrations.

Based on the configurations obtained from convex hull, we can compute the average intercalation voltage as a function of concentration. For each stable M$_x$VO$_2$ compound, the voltage (V) is calculated as:

$$V = -\frac{E(x_2) - E(x_1) - (x_2 - x_1)E(M^{bulk})}{z(x_2 - x_1)} \quad (2)$$

where $E(x_2)$ and $E(x_1)$ are the total energies of the M$_x$VO$_2$ compound at two neighbouring low-energy concentrations $x_2$ and $x_1$ along the computed convex hull (solid line in Fig. 3b), and $E(M^{bulk})$ is the energy per atom of M=Li, Mg or Al in the bulk form, z is the valence charge.

Fig. 3c shows the calculated voltage-composition curve of Li$_x$VO$_2$. The shape of the calculated voltage curve is similar to the experimental one,[38] though some discrepancy exists in the middle region. As Li concentration in Li$_x$VO$_2$ increases, the electrostatic Li-Li repulsion becomes more significant, leading to the drop in voltage. At each metal-ion concentration, the calculated Li voltage is consistently larger than Mg and Al voltages. Despite the lower voltage, the energy density of Mg-ion and Al-ion batteries would be still competitive due to the ability of Mg and Al to exchange two and three redox electrons per cation, respectively. The calculated voltage for the full lithiation of VO$_2$ to LiVO$_2$ compound is 2.23 V, close to the experimental value of ~2.0 V in rutile VO$_2$ nanowires.[38]

As shown in Fig. 3c, the computed voltage for Mg insertion initially starts at 2.10 V, followed by a plateau at 1.62 V up to Mg$_{0.5}$VO$_2$ and then a (two-step) plateau at ~0.50 V up to Mg$_1$VO$_2$. The combination of theoretical capacity of 323 mAh/g and voltage of 1.6 V in VO$_2$(R) would result in specific energies greater than those of the state-of-the-art Mg cathodes, such as TiS$_2$ (190 mAh/g at 1.1 V),[64] α-MnO$_2$ (280 mAh/g at 0.87-1.5 V)[65] and V$_2$MoO$_8$ (312 mAh/g at 1.32 V).[66]

Fig. 3c shows the calculated voltage-composition curve of Al$_x$VO$_2$. The voltage profile has multiple plateaus, beginning at about 1.98 V and dropping to about 1.48 V at Al$_{0.25}$VO$_2$, followed by a plateau at around 1.17 V. The computed specific capacity for Al insertion in rutile VO$_2$ (480 mAh/g at 1.17 V) is among the highest for existing cathode materials, such as V$_2$O$_5$ (200mAh/g at ~1 V),[67] SnS$_2$ (392 mAh/g at 0.68 V)[68] and CuS@C nanocomposite (240 mAh/g at 0.6 V).[69] Overall, the electrochemical properties predicted by our DFT+U calculations suggest that VO$_2$(R) is a very promising cathode material for multivalent batteries.

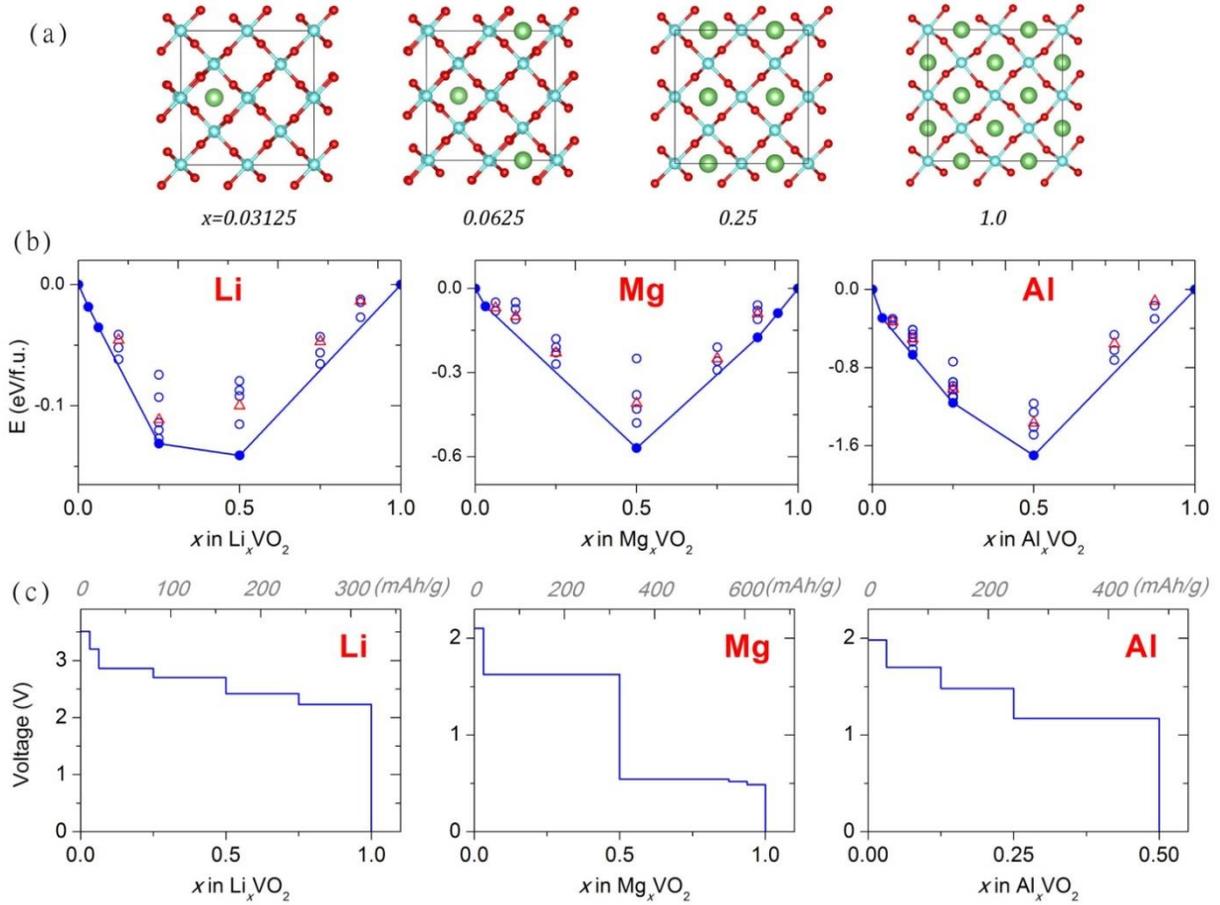

**Fig. 3** (a) Selected optimized structures of $Li_xVO_2$. The V, O and Li atoms are shown in cyan, red and green colors, respectively. (b) The convex hull as a function of Li, Mg and Al content in $VO_2(R)$. Circles and triangles represent metal ions occupying octahedral and tetrahedral sites, respectively. (c) Calculated voltage profiles for Li, Mg and Al insertion in $VO_2(R)$

### 3.4. Volume changes

Next, we tested the structural stability of the $VO_2(R)$ host during Li, Mg, and Al insertion. The extent of volume changes is an important factor for battery performance especially in the materials with multi-phase reactions since those reactions induce additional strain at the phase boundaries. Shin et al demonstrated that the amount of capacity loss is proportional to the difference in lattice parameters between the phases.[70] Fig. 4a shows the variation of the lattice parameters of $VO_2(R)$ during Li intercalation. We find that Li insertion leads to the 8-9% increase in the $a$ and $b$ lattice constants from $VO_2$ to $LiVO_2$. In contrast, the $c$ parameter increases only slightly (3 %), followed by a 2.5% contraction after $Li_{0.25}VO_2$. Such anisotropic behaviour in $VO_2(R)$ qualitatively agrees very well with the experimental data on other rutile materials, such as $MnO_2$.[71] For instance, Jiao et al. reported that full lithiation of mesoporous $MnO_2$ led to 13.9% expansion along the $a$ direction and a 2% contraction along the $c$ direction.[71] The volume of $VO_2$ supercell increases linearly with Li, Mg and Al concentration by 17.6% to $LiVO_2$, 34.6% to $MgVO_2$ and 26.4% to $Al_{0.5}VO_2$. The volume expansion during Li insertion compares well to other rutile oxides (15% in $LiTiO_2$[72] and 25-30% in $LiMnO_2$[71, 73]).

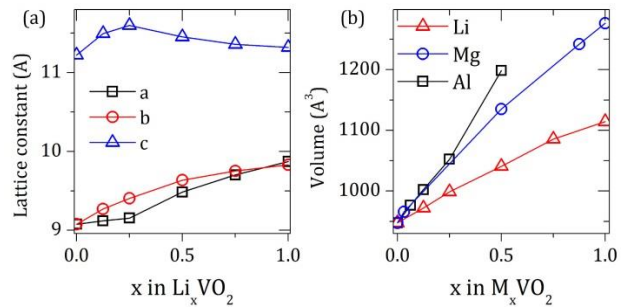

**Fig. 4** The change of (a) lattice parameters and (b) volume for different compositions of $M_xVO_2$ (M = Li, Mg, Al)

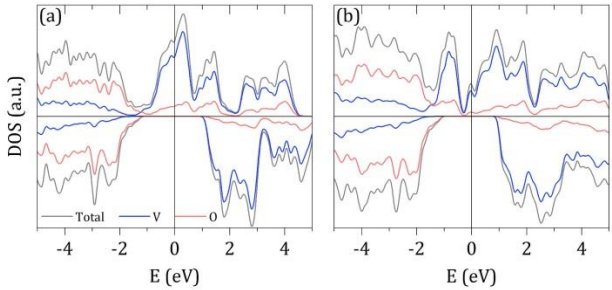

**Fig. 5** Projected spin-polarized DOS of (a) pristine $VO_2$, and (b) Li-inserted $VO_2$. The Fermi level is aligned to zero, indicated by the vertical line

### 3.5. Electronic structure and charge transfer mechanism

Rutile $VO_2$ phase has spin-polarized metallic electronic structure with large density of states at the Fermi level, as shown in Fig. 5a. This is in good agreement with the experimental conductivity measurements[74] and other theoretical studies.[59] The states near the Fermi level originate mainly from V 3$d$ states. The $p$–$d$ hybridization causes additional O 2$p$ contributions in this energy range as well. We find that Li, Mg, Al insertion does not significantly affect the electronic structure of the $VO_2$ host, as show in Fig. 5b on the example of Li intercalation.

The detailed mechanism of Li, Mg and Al insertion was investigated based on the charge redistribution. The Bader charge analysis shows that Li, Mg and Al atoms are almost completely ionized and donate up to 0.87, 1.61 and 2.45 |e|, respectively. Consequently, we found a charge accumulation on the V atoms, indicating the reduction of V species from $V^{+4}$ to $V^{+3}$. Interestingly, we can notice that a large fraction of the donated charge is not only transferred to V, but also to oxygen. Similar observation has also been reported for $LiCoO_2$ system.[62] The Bader charge analysis shows that the electron transfer from Li, Mg or Al ions to the $VO_2$ host is very local. In particular, there is a significant increase in the electron density at the V and O atoms in the first coordination sphere of the inserted Li, Mg or Al. The contribution from oxygen reduction is important as it allows maintaining positive voltages e.g. in the case of Al up to $Al_{0.5}VO_2$ and not up to $Al_{0.33}VO_2$ which would follow from only one-electron vanadium reduction.

### 3.6. Li, Mg and Al diffusion

The charge/discharge rate capability of the electrode material depends on the kinetics of electron transport and ionic diffusion. The previous analysis showed that $VO_2(R)$ has a metallic electronic structure with large density of states at the Fermi level. Hence, the electron conductivity in the $VO_2(R)$ anode is expected to be sufficiently high. Examination of Li, Mg and Al mobility and diffusion pathways thus becomes of crucial importance. Achieving fast ionic diffusion still remains a challenge in practical multivalent batteries.

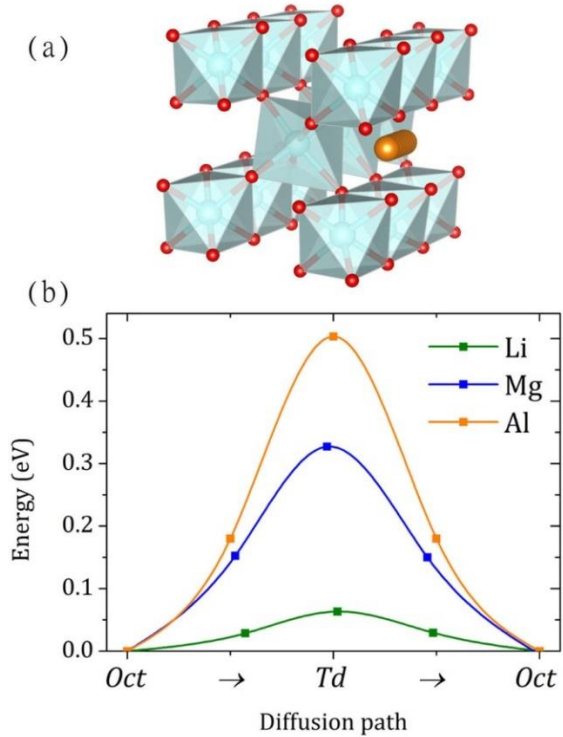

**Fig. 6** (a) Diffusion pathway (Oct-Td-Oct), and (b) corresponding energy profile for Li, Mg and Al diffusion in $VO_2(R)$

There are few possible diffusion pathways in the rutile structure, corresponding to: (1) movement in the [010] or [111] direction directly through the edge-sharing VO6 octahedrons (*inter-tunnel diffusion*), and (2) movement within the open channel along the *c*-axis in the [001] direction (*intra-tunnel diffusion*). However, previous studies on rutile $TiO_2$ and $RuO_2$ demonstrated that the barrier for inter-tunnel Li hopping can be as high as 3-7 eV, making this diffusion pathway highly unlikely.[72, 75] The crystal structure of the rutile host, therefore, favours Li, Mg and Al transport along the 1D channels.

Figure 6a shows the typical diffusion pathway between two neighbouring Oct sites along the c-axis in $VO_2(R)$. The diffusion occurs in the form of jumps between the neighbouring *Oct* sites, while the *Td* position is a transition state. The corresponding energy profile as calculated by the NEB method[45] is shown in Fig. 6b. We find that the energy barrier for Li diffusion via the *Oct*→*Td*→*Oct* pathway in $VO_2(R)$ is 0.06 eV. The Li migration barriers are expectedly low, in good agreement with experimentally observed fast Li intercalation in heteroepitaxy-grown $VO_2$.[38]

The calculated energy barriers for Mg and Al diffusion in $VO_2(R)$ are only 0.33 and 0.50 eV, respectively. The relative order of activation energies (Li<Mg<Al) is consistent with the stronger interactions between a multivalent intercalant and the host environment. The calculated barriers for Mg and Al diffusion are among the lowest reported for intercalation materials.[76-78] The Mg and Al diffusion barriers are lower than in other VO oxides, such as $V_2O_5$ (0.65 eV), which can be attributed to the smaller coordination change between minimum-energy and saddle points during the migration process (6→4→6 in $VO_2$ versus 8→3→8 in α-$V_2O_5$). The lower oxidation state of vanadium in $VO_2$ versus $V_2O_5$ is beneficial as well due to the weaker electrostatic interaction with the migrating cation.[79] The relatively low diffusion barriers in $VO_2(R)$ are consistent with small differences in total energy between the Oct and Td sites (0.08, 0.14 and 0.23 eV for Li, Mg and Al insertion, respectively). In comparison, the Oct-Td energy difference in rutile $TiO_2$ is as large as 0.7 eV for Li insertion.[72] As demonstrated by Liu et al,[80] there is often a strong correlation between the site energy difference for a cation and corresponding activation energy for diffusion. The calculated barrier for Al diffusion in $VO_2(R)$ is one of the lowest among all electrode materials for Al-ion batteries reported in the literature. Although it is higher than the energy barrier in expanded graphite (0.02-0.33 eV),[81, 82] but it is comparable to $BC_3$ electrode (0.38-1.20 eV)[83] and lower than in any oxide framework reported up to date.

Based on the calculated energy barriers, it is possible to estimate the Li, Mg and Al diffusivity in $VO_2(R)$. Under the assumption of only nearest-neighbour hops among equivalent sites, the diffusion constant for a 1-D tunnel takes a simple form of $D = \Gamma a^2$, where $\Gamma$ is the hopping frequency and $a$ is the hopping length.[61, 84] Based on a transition state theory, the diffusion constant can be estimated as

$$D = a^2 v^* exp(-E_b/k_B T) \quad (3)$$

where $v^*$ is the attempt frequency, $k_B$ is a Boltzmann constant and $T$ is temperature. As shown by Rong et al,[85] the $E_b$ of ~0.52 eV corresponds to a room-temperature ionic diffusivity of $10^{-12}$ cm$^2$ s$^{-1}$. Moreover, each 0.06 eV increase in the migration barrier leads to an order of magnitude decrease in diffusivity.[86] Considering the relationship between diffusivity and particle size, migration barriers up to 0.65 eV were suggested as guideline for adequate battery operation of nano-sized particles.[85, 87] Importantly, our calculated activation energies (0.06, 0.33 and 0.50 eV for Li, Mg and Al, respectively) fit within the above criteria very well. These results suggest that tunnel structure of rutile $VO_2$ can provide fast ionic transport, resulting in a high specific capacity even at a high charge/discharge current. To make use of this fast transport, rational experimental design of $VO_2(R)$ with abundant (002) surfaces for metal-ion insertion would be desirable.[38]

## Conclusions

Using first-principles calculations, we systematically investigated thermodynamic, electronic and kinetic properties associated with the insertion of Li, Mg and Al atoms in rutile $VO_2$ We demonstrated the excellent Li, Mg and Al mobility and high structural stability in $VO_2$. The calculated Li intercalation voltages are from 3.43 V to 2.23 V for $Li_xVO_2$ for $x$=0.03 to 1, which agrees well with the experimental results. The computed voltage for Mg insertion is about 1.6 V up to $Mg_{0.5}VO_2$ followed by a plateau at about 0.5 V up to $Mg_1VO_2$, which is quite promising. The predicted voltage curve for Al insertion has multiple plateaus, beginning at about 1.98 V and dropping to 1.48 V at $Al_{0.25}VO_2$, followed by a plateau at 1.17 V. Predicted Al voltage is competitive with other proposed cathode materials such as $V_2O_5$ and $AlCl_3$/graphite. The diffusion barrier of Li, Mg and Al in the tunnel structure of $VO_2$ is 0.06, 0.33 and 0.50 eV, respectively. These values are very much competitive over the existing cathode materials and are compatible with high-rate operation. The analysis of electronic structure suggests that the mechanism of Li, Mg and Al insertion significantly relies on contributions from oxygen redox. Charge accommodation by oxygen atoms enables achieving higher capacities than what would follow based on the simple rule of 1 valence electron being accommodated by 1 vanadium atom.

Overall, our theoretical work fits very well with the latest experimental advances in controlled growth and epitaxial stabilization of the VO polymorphs as reported recently.[38, 88] Heteroepitaxy-controlled and specifically-oriented growth can be used for activating the anisotropic rutile structure and improving Li diffusion kinetics.[38, 88] These results not only help to understand the intrinsic mechanism of the phenomenon observed in the Li-ion battery experiments, but also suggest significant potential rutile VO$_2$ electrodes for multivalent batteries.

**Acknowledgements**

This work is supported by the Ministry of Education of Singapore grant MOE2015-T2-1-011.

**Supporting information**

**Table S1.** Relative total energies (in meV per formula unit) of FM and AFM $VO_2$ and Li-$VO_2$. Zero energy is referenced to FM state

|     | $VO_2$ | $Li_{0.33}VO_2$ | $Li_{0.5}VO_2$ | $LiVO_2$ |
|-----|--------|-----------------|----------------|----------|
| FM  | 0      | 0               | 0              | 0        |
| AFM | +45    | +43             | +53            | +52      |

**Table S2.** Voltage for Li insertion in rutile and monoclinic phases of $VO_2$

| Composition      | $V$ (V) (rutile) | $V$ (V) (monoclinic) |
|------------------|------------------|----------------------|
| $Li_{0.03}VO_2$  | -3.50            | -3.27                |
| $Li_{0.25}VO_2$  | -2.81            | -2.59                |